\documentclass[10pt, final, journal]{IEEEtran}

\usepackage{fancyhdr}
\usepackage{hyperref}
\usepackage{cite}
\usepackage{graphicx}
\usepackage{amsmath}
\usepackage{amssymb}
\usepackage{bm}
\usepackage{times}
\usepackage{color}
\usepackage{url}
\usepackage{array}
\usepackage{float}
\usepackage{caption}
\usepackage{subcaption}
\usepackage{multirow}
\usepackage{rotating}
\usepackage{extarrows}
\usepackage{multicol}
\usepackage{epstopdf}
\usepackage{tikz}
\usetikzlibrary{arrows,shapes}
\tikzset{
 state/.style={
            rectangle,
            draw=black, 
            minimum height=2em,
            inner sep=2pt,
            text centered,
            }
}

\def\wh{\widehat}
\def\ex{\mathbb{E}}
\def\mb{\mathbf}
\def\mc{\mathcal}

\def\shrnk{\hspace{-2px}}
\newcommand{\partl}[2][]{\frac{\partial#1}{\partial#2}}
\newcommand{\der}[2][]{\frac{d#1}{d#2}}
\newtheorem{thm}{Theorem}
\newtheorem{prpn}[thm]{Proposition}

\begin{document}
\title{\huge Robust Energy Harvesting Based on a Stackelberg Game}
\author{\IEEEauthorblockN{Siddhartha Sarma, Kundan Kandhway and Joy Kuri}
	\thanks{The authors are with the Department of Electronic Systems Engineering, Indian Institute of Science, Bangalore, Karnataka - 560012, India (e-mail: \{siddharth, kundan, kuri\}@dese.iisc.ernet.in).} }

\maketitle
\thispagestyle{fancy}

\begin{abstract}
We study a Stackelberg game between a base station and a multi-antenna power beacon for wireless energy harvesting in a multiple sensor node scenario. Assuming imperfect CSI between the sensor nodes and the power beacon, we propose a utility function that is based on \emph{throughput non-outage probability} at the base station. We provide an analytical solution for the equilibrium in case of a single sensor node. For the general case consisting of multiple sensor nodes, we provide upper and lower bounds on the power and price (players' strategies). We compare the bounds with solutions resulting from an exhaustive search and a relaxed semidefinite program, and find the upper bound to be tight.
\end{abstract}
\begin{IEEEkeywords}
	Wireless energy harvesting, Stackelberg game, Imperfect CSI, Beamforming, Sensor networks
\end{IEEEkeywords}
\vspace{-0.5cm}
\section{Introduction}
Recently, due to the ubiquitous presence of wireless devices, wireless power transfer (WPT)\cite{varshneyISIT2008,DiamantoulakisArxiv2015} has grabbed the attention of researchers. WPT has huge application in scenarios such as sensor networks in a border area or a toxic zone, where regular replacement of batteries is almost impossible. Traditionally, wireless sensor nodes rely on batteries with limited lifetime to transmit data to a base station. In a wireless power harvesting scenario, along with the base station, power beacons are also deployed to recharge the sensor nodes. In the absence of any internal power source, these sensor nodes follow a \textit{harvest-then-transmit} \cite{juTWC2014} scheme to communicate with the base station.

In practice, sensor networks and power beacons may be deployed by different authorities. As a result, one needs a mutually beneficial scheme that ensures \emph{energy trading} between both parties. Such scenarios involving multiple self-interested agents can be studied in a game theoretic framework. The authors in \cite{chenICASSP15} have studied energy trading between power beacons and their users by formulating a Stackelberg game.

Our modeling approach and analysis are different from those in \cite{chenICASSP15}. Unlike \cite{chenICASSP15}, where single antenna power beacons are considered, we consider a \emph{multi-antenna power beacon} \cite{xingSPL2013,LiuTC2014,WuWCL2016}. Such multi-antenna systems can improve the efficiency of energy transfer by employing beamforming. Contrarily, a base station with a single antenna can cater to the requirements of low data rate sensor networks and is considered in our work.

In \cite{chenICASSP15}, perfect channel state information (CSI) between the power beacons and the sensor node is assumed. But, in real-world scenarios, one requires training signals to estimate channel gains and a higher accuracy can only be obtained at the cost of a higher estimation time. For multi-antenna systems, the issue is more critical compared to single antenna as efficient beamforming requires accurate CSI. Also, unlike capacity, a logarithmic (sub-linear) function of channel gain, harvested energy is a quadratic (super-linear) function of channel gain and, therefore, inaccurate CSI has further degrading effects.
\begin{figure}[!t]
\vspace{-0.3cm}
	\centering
	\includegraphics[scale=0.4]{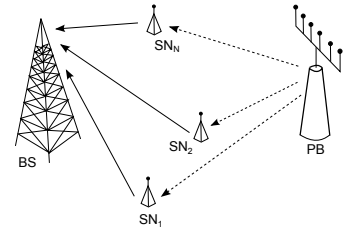}
	\caption{\small Graphical representation of energy harvesting sensor network.}
	\label{fig:model}
\vspace{-0.7cm}
\end{figure}
Therefore, considering imperfect CSI (e.g., based only on path-loss) for the power beacon---which has multiple antennas---is more realistic, and this is considered in our work. Also, we generalize the model by considering multiple sensor nodes. 
 
Due to imperfect CSI \cite{xingSPL2013}, our problem formulation differs from that in \cite{chenICASSP15} significantly. In \cite{chenICASSP15}, utilities are deterministic, whereas, in our formulation the base station's utility is based on a \emph{probabilistic} term in the proposed Stackelberg game. As we cannot directly calculate the data throughput, we consider a utility function that is based on \emph{throughput non-outage probability}---probability of throughput being above a predefined threshold. Outage probability of an adhoc network with wireless power transfer was evaluated in \cite{guoWCL15} in a non-game theoretic setting and significantly differs from ours. Also, we differ from \cite{juTWC2014,xingSPL2013} as in those works, the sensor network and power beacon belonged to the same deploying authority.

Our work ensures quality of sensing by keeping the data throughput for each sensor node above a certain threshold. Outage occurs when the throughput goes below that threshold. Therefore, we would like to improve the minimum non-outage probability over the sensor nodes (equivalent to decreasing the maximum outage probability) by asking the power beacon to adjust the antenna gains appropriately while transmitting sufficient power. The base station should compensate the power beacon monetarily for its service. So, the utility of the base station should include the revenue generated due to non-outage of data throughput and compensation paid to the power beacon (with appropriate scaling). The utility of the power beacon should consist of the revenue generated by selling power, minus the operational cost.

Due to intractability of the throughput non-outage probability, we bound it using Markov's and Jensen's inequalities.  We then formulate a Stackelberg game with these utility functions whose equilibrium simultaneously maximizes the respective utilities. The solution of this Stackelberg game decides the optimal antenna weights and transmit power for the power beacon for single sensor node scenario. For the multiple sensor scenario, we provide computationally efficient solutions by bounding the base station's utility function.

The main contributions with respect to prior works (including \cite{chenICASSP15}) are: (i) Formulation of a Stackelberg game for multiple sensor nodes with a multiple antenna power beacon assuming imperfect CSI between the sensor nodes and the power beacon, (ii) Analytical evaluation of the optimal solution for the Stackelberg game  for a single sensor node, (iii) The equilibrium strategies (power and price) of the players corresponding to the upper and lower bounds on the base station's utility for the most general case involving multiple sensor nodes, (iv) Comparison among the solutions provided by the bounds, a relaxed semidefinite program and exhaustive search.

\section{System Model \& Problem formulation}
The system model consists of one base station (BS), one multi-antenna power beacon (PB) and $N$ sensor nodes (SN), indexed by $i \in  \mc{N}= \{1,2,\cdots,N\}$ as shown in Fig. \ref{fig:model}. The power beacon has $M$ antennas, whereas, the base station and sensors are equipped with a single antenna each. All sensor nodes can communicate directly with the base station. We adopt the \textit{harvest-then-transmit} model \cite{juTWC2014}. We consider a slotted system in which each time slot is divided into two parts: (i) In the first part, which consists of a fraction  $\tau$ of the total time slot, the power beacon transmits power to the sensor nodes for energy harvesting; (ii) In the second part (of duration $(1-\tau)$ fraction of the time slot), the sensor nodes transmit data (information) to the base station using the harvested energy. In several scenarios, different authorities may deploy the sensor network and the power beacons; so, the base station needs to negotiate with the power beacon for wireless power transfer. In the proposed Stackelberg game, the power beacon acts as a seller and charges the base station at the rate $\rho$ per unit power and as a consequence, the base station asks for $P$ units of power from the power beacon.

Without loss of generality, we can assume that each time slot is of unit length. During the first part, energy harvested at the $i$th sensor node is given by:
\begin{align}\label{eq:harvest}
	E_i=\tau|\mb{h}_i^\dagger\mb{w}|^2P, \quad \text{where } \mb{w^\dagger w}=1
\end{align} 
where $\mb{h}_i$ is the $M\times1$ complex channel gain vector from the power beacon to the $i$th sensor node, and $\mb{w}$ is the complex antenna gain vector of the same dimension. The second part of the time slot is divided into $N$ equal parts.\footnote{One can adjust $\tau$ and transmission times assigned to the $N$ sensor nodes based on requirements. Many TDMA systems \cite{demirkolCM2006} use predefined time slots for operational simplicity. We have assumed equal time slots as an example.} Each sensor node transmits data to the base station in its assigned time slot by completely spending the energy harvested during the first part. The data throughput at the base station due to the $i$th sensor node can be written as:
\begin{align*}
D_i(P,\mb{w})=\frac{(1-\tau)}{N}\left[ \frac{1}{2}\log_2\left( 1+|h_{si}|^2\frac{E_i}{\sigma^2((1-\tau)/N)}\right) \right].
\end{align*}
Using \eqref{eq:harvest}, we get
\begin{align}
D_i(P,\mb{w})= \frac{(1-\tau)}{2N}\log_2\left( 1+\frac{\tau N|h_{si}|^2|\mb{h}_i^\dagger\mb{w}|^2P}{\sigma^2(1-\tau)}\right) \label{eq:thrput}.
\end{align}
Here, $h_{si}$ is the complex channel gain between the $i$th sensor and the base station. In the throughput equation, we have assumed that the Gaussian noise is characterized by zero mean and variance $\sigma^2$. We assume that the base station has perfect information about all $h_{si}$ (can be obtained using training signals), whereas, the base station has imperfect information about all $\mb{h}_i$ (as mentioned before).
Due to imperfect CSI, throughput cannot be calculated; therefore, the base station would like to maximize \emph{throughput non-outage probability} given by:
\begin{align}\label{eq:nonoutgeprob}
P_{non-outage,i}(\beta) = \Pr\left(D_i(P,\mb{w})>\beta\right).
\end{align}
Here, $\beta>0$ is a predefined threshold. To ensure quality of sensing, the base station would try to improve the minimum $P_{non-outage,i}(\beta)$ over the sensor nodes by purchasing appropriate amount of power and by asking the power beacon to adjust the antenna gains. However, at the same time, expenditure on power should be reduced, leading to a weighted metric.

We model the imperfect CSI between the power beacon and the $i$th sensor node in the following manner: $\mb{h}_i=\mb{\wh{h}}_i+\mb{e}_i$, where $\mb{\wh{h}}_i$ is the estimated channel and the error $\mb{e}_i$ is a random vector, which is distributed according to circularly symmetric complex Gaussian distribution, i.e., $\mathcal{CN}(0,\mathbf{\Sigma})$, where $\mathbf{\Sigma}$ is  the covariance matrix. Consequently, from \eqref{eq:thrput}, $P_{non-outage,i}(\beta)$ is a logarithmic function of a nonlinear combination of multiple Gaussian random variables. A closed form expression for $P_{non-outage,i}(\beta)$ is difficult to obtain, therefore we propose the following bound on $P_{non-outage,i}(\beta)$ by applying the Markov's and Jensen's inequalities.

\vspace*{-0.3cm}
\small
\begin{align}\label{eq:outage}
	\hspace*{-5pt}
	\Pr\big( & D_i(P,\mb{w}) \ge \beta \big)  \le \frac{\ex{\left[ D_i(P,\mb{w})\right]}}{\beta} \nonumber \\
	&  \le \frac{(1-\tau)}{2\beta N}\log_2\left( 1+\frac{\tau|h_{si}|^2\ex{\left[ |\mb{h}_i^\dagger\mb{w}|^2\right] }P}{(1-\tau)\sigma^2/N}\right) \nonumber \\
	& = \frac{(1-\tau)}{2\beta N}\log_2\left( 1+\frac{\tau|h_{si}|^2N\mb{w}^\dagger\ex{\left[ \mb{h}_i\mb{h}_i^\dagger\right]}\mb{w}P}{(1-\tau)\sigma^2}\right)
\end{align}

\normalsize
Denoting the right hand side of \eqref{eq:outage} by $\Gamma_i(P,\mathbf{w})$, the utility of the base station is:
\begin{align*} 
U_{BS}(\rho,P,\mb{w})=\min_i\;\lbrace\alpha \Gamma_i(P,\mathbf{w})\rbrace-\tau\rho P,
\end{align*} 
where $\alpha (>0)$ is a weighting parameter. The utility maximization problem for the base station is:
\begin{align}\label{opt:bs}
	\max\; U_{BS}(\rho,P,\mb{w}),\:\:
	\text{subject to: } P \ge 0,\quad
	\mb{w}^\dagger\mb{w}=1
\end{align}

We define the following utility function for the power beacon: $U_{PB}(\rho,P)=(\rho-c)P$, where $c$ captures the operational cost per unit transmitted power. Note that if $\rho<c$, then the utility is negative and power beacon would refuse to sell power. The relevant optimization problem at the power beacon is:
\vspace{-1.5em}
\begin{align}\label{eq:PButil}
\max\;U_{PB}(\rho,P)\:\: \text{subject to: }\rho \ge c.
\end{align}
We solve the problems \eqref{opt:bs} and \eqref{eq:PButil} to obtain the optimal antenna weight $\mb{w}^*$, optimal price $\rho^*$ and optimal power $P^*$ that simultaneously maximize the utilities of both the players.
\section{Analysis \& Solution}
In Sec \ref{sec.singsensenode}, we consider a scenario containing a single sensor node and obtain a closed form solution for the proposed Stackelberg game. In Sec \ref{sec.multsensnode}, we address the multiple sensor node scenario.
\vspace{-0.5cm}
\subsection{Single sensor node (N=1)}\label{sec.singsensenode}
For the single sensor node scenario, $\Gamma=(1-\tau)\ln\left(1+\mu\kappa P\right)/(2\beta \ln 2)$, where $\kappa=\tau|h_s|^2/(\sigma^2(1-\tau))$, $\mu=\mb{w}^\dagger\mb{Q}\mb{w}$ and $\mb{Q}=\ex{\left[ \mb{h}\mb{h}^\dagger\right]}=\wh{\mb{h}}\wh{\mb{h}}^\dagger+\mathbf{\Sigma}$. The utility function of the base station for this scenario can be rewritten as:
\begin{align}\label{eq:main_BS_util}
U_{BS}(\rho,P)=\alpha'\ln\left(1+\mu\kappa P\right)- \tau\rho P,
\end{align}
where, $\alpha'=\alpha(1-\tau)/(2\beta \ln 2)$. 

Since, we are considering normalized antenna weights and the optimal antenna weight is 
independent of $P$, therefore $\mb{w}^\dagger\mb{Q}\mb{w}$ can be maximized in isolation using the following optimization problem:
\begin{align}\label{opt:weight}
\mu^*= \max_\mb{w}\; \mb{w}^\dagger\mb{Q}\mb{w}\:\:\text{subject to: }\mb{w^\dagger w}=1
\end{align}

Optimization problem \eqref{opt:weight} is the \emph{Rayleigh Quotient problem} \cite[p.~176]{Horn1985} and, therefore, the optimal objective function value, $\mu^*$ and the solution, $\mb{w}^*$ are the maximum eigenvalue of the matrix $\mb{Q}$ and the corresponding eigenvector, respectively. 

First, we optimize the base station's utility for a fixed $\rho$. Differentiating the base station's utility function with respect to $P$, we get:

\small
\vspace{-0.3cm}
\begin{align} 
\partl[U_{BS}]{P} =&\alpha'\frac{\mu^*\kappa}{(1+\mu^*\kappa P)}-\tau\rho=0 \nonumber\\ 
\implies & P^*(\rho)=\frac{1}{\mu^*\kappa}\left( \frac{\alpha'\mu^*\kappa}{\tau\rho}-1\right) \label{eq:power}
\end{align}
\normalsize

Second, we replace the value of $P$ (from \eqref{eq:power}) in the seller's utility function \eqref{eq:PButil} and maximize with respect to $\rho$. 

\small
\begin{align}
\der[U_{PB}]{\rho} & = P +(\rho-c)\der[P]{\rho}=0 \nonumber\\  \implies & \frac{1}{\mu^*\kappa}\left( \frac{\alpha'\mu^*\kappa}{\tau\rho}-1\right) -(\rho-c)\frac{\alpha'}{\tau\rho^2}=0 \nonumber \\
\implies & \rho^* =\sqrt{\alpha'c\mu^*\kappa/\tau} \label{eq:optrho}
\end{align}
\normalsize
Substituting \eqref{eq:optrho} in  \eqref{eq:power}, we calculate the optimal power,

\small
\begin{align}\label{eq:optpow}
P^*=\frac{1}{\mu^*\kappa}\left( \sqrt{\frac{\alpha'\mu^*\kappa}{\tau c}}-1\right)
\end{align} 
\normalsize

Note that as both $U_{BS}$ and $U_{PB}$ are concave functions, so $P^*$ and $\rho^*$ are equilibrium points of the proposed Stackelberg game.

\textit{Special case (M=1):} For this single antenna power beacon and single sensor node case, we have an alternate approach for calculating the equilibrium. Note that we only need to find the optimal power value (beamforming is absent). Rearranging the terms in the probability expression in Eq. \eqref{eq:nonoutgeprob}:
\begin{align}\label{eq:singlantenna}
	P_{non-outage}(\beta)=\Pr\left(\shrnk|h|^2\shrnk \ge \shrnk \frac{(4^\frac{\beta}{(1-\tau)}-1)(1-\tau)\sigma^2}{\tau|h_s|^2P} \right)
\end{align}
$h=\wh{h}+e$, where the error $e$ is distributed according to $\mathcal{CN}(0,\sigma_1^2)$. Thus, $|h|^2$ is   a non-central chi-square random variable with degree of freedom (d.o.f.) $2$ and non-centrality parameter  $\theta^2=2\left( \frac{\Re(\wh{h})}{\sigma_1}\right) ^2+  2\left( \frac{\Im(\wh{h})}{\sigma_1}\right) ^2=  \frac{2|\wh{h}|^2}{\sigma_1^2}$. Further, we can approximate the non-central chi square distribution of d.o.f. $2$ by a central chi-square random variable ($\chi_0^2$) of the same d.o.f. \cite{coxCJS1987}. Recalling that the chi square random variable with d.o.f $2$ is an exponential random variable, from Eq. \eqref{eq:singlantenna}, we can write

\begin{align*}
	P_{non-outage}(\beta)\shrnk\approx\shrnk \Pr\left(\chi_0^2 \ge  \frac{\eta}{P} \right) \shrnk=\shrnk\exp\left(-\frac{\eta}{P}\right),
\end{align*}
where, 
${\eta= \frac{(4^\frac{\beta}{(1-\tau)}-1)(1-\tau)\sigma^2}{\tau|h_s|^2(1+\theta^2/2)}}$.

The approximation considered above is quite accurate for smaller values of centrality parameter ($\theta^2 < 0.4$). Now, the utility can be written as:
\begin{align}\label{eq:utilbssingle}
	U_{BS}(P)=\alpha'\exp(-\eta/P)-\tau\rho P 
\end{align}
\begin{prpn}\label{prpn:chi2}
	The optimal solution can be calculated by solving the following system of equations in $P$ and $\rho$:
	\begin{align}\label{eq:prpnchi2}
		\alpha'\eta e^{-(\eta/P)}-\tau\rho P^2 =0 \textrm{ and }\eta/P-c/\rho  =1.
	\end{align}
\end{prpn}
\begin{IEEEproof}
	Differentiating Eq. \eqref{eq:utilbssingle} with respect to $P$ and equating it to $0$, we get:
	\begin{align}\label{eq:diffexctoutg}
	\der[U_{BS}]{P}=\frac{\alpha'\eta\exp(-\eta/P)}{P^2}-\tau\rho=0
	\end{align}
	Similarly, for utility of the power beacon, we get:
	\begin{align}\label{eq:diffpbutil}
	\der[U_{PB}]{\rho}=P+(\rho-c)\der[P]{\rho}=0
	\end{align} 
	For a given $\rho$, Eq. \eqref{eq:diffexctoutg} provides us the optimal utility. Differentiating it with respect $\rho$, we get (using Eq. \eqref{eq:diffexctoutg}):
	\begin{align*}
	\left( \frac{\alpha'\eta^2\exp(-\eta/P)}{P^2}-2\tau\rho P\right) \der[P]{\rho}=\tau P^2	\Rightarrow \rho\left( \frac{\eta}{2P}-1\right)\der[P]{\rho}=\frac{P}{2}
	\end{align*}
	Replacing this in Eq. \eqref{eq:diffpbutil}, we get the second equation, whereas, the first equation corresponds to Eq. \eqref{eq:diffexctoutg}.
\end{IEEEproof}

\subsection{Multiple sensor nodes}\label{sec.multsensnode}
For multiple sensor nodes, the base station utility is:
\footnotesize
\begin{align}\label{eq:multisensorbsutil}
{
U_{BS}(\rho,P,\mb{w})=\min_i\;\frac{\alpha(1-\tau)}{2N\beta}\log_2\left( 1+\frac{N\tau|h_{si}|^2\mb{w}^\dagger\mb{Q}_i\mb{w}P}{(1-\tau)\sigma^2}\right) -\tau\rho P.}
\end{align}

\normalsize
Here $\mb{Q}_i=\ex{[\mb{h}_i\mb{h}_i^\dagger]}$ and $h_{si}$ is the $i$th sensor-to-base station channel gain.

As in the previous section, we can optimize antenna gain separately to maximize utility. We formulate the following optimization problem to obtain the optimal $\mb{w}$.
\begin{subequations}\label{opt:wght_opt}
	\begin{align}
	\max\;\; & \nu \\
	\text{subject to: } & |h_{si}|^2\mb{w}^\dagger\mb{Q}_i\mb{w} \ge \nu,\:\: \forall i\in \mc{N},\:\:  \mb{w^\dagger w}=1,
	\end{align}
\end{subequations}

Problem \eqref{opt:wght_opt} is non-convex, so a global solution is difficult to obtain. But we can obtain analytical upper and lower bounds efficiently. Let
\vspace{-0.3cm}
\begin{subequations}\label{eq:bounds}
	\begin{align}
	\nu_{min}\shrnk=\shrnk\min\{\lambda_{min}(|h_{si}|^2\mb{Q}_i),\:\forall i\},\label{eq:lowbnds} \\
	\nu_{max}\shrnk=\shrnk\min\{\lambda_{max}(|h_{si}|^2\mb{Q}_i),\:\forall i\} \label{eq:upbnds}
	\end{align}
\end{subequations}

\vspace{-0.05cm}
Here, $\lambda_{min}(.)$ and $\lambda_{max}(.)$ denote the minimum and maximum eigenvalue of a matrix. Using the value of $\nu_{min}$ or $\nu_{max}$ (instead of $\mu^*/|h_s|^2$) in \eqref{eq:optrho} and \eqref{eq:optpow}, we can solve the Stackelberg game for these two values.

Alternatively, using $\mb{w}^\dagger\mb{Q}_i\mb{w}\hspace{0pt}=\hspace{0pt}Tr(\mb{w}^\dagger\mb{Q}_i\mb{w})\hspace{0pt} =\hspace{0pt}Tr(\mb{Q}_i\mb{w}\mb{w}^\dagger)\hspace{0pt}$
$ =\hspace{0pt}Tr(\mb{Q}_i\mb{W})$, and relaxing the rank-1 constraint $\mb{W}=\mb{ww^\dagger}$, Problem \eqref{opt:wght_opt} results in the following relaxed semidefinite program:

\vspace{-0.4cm}
\small
\begin{subequations}\label{opt:sdp}
\begin{align}
	 \max&\:\:\nu \\
 \text{s.t.\::}&\:Tr(|h_{si}|^2\mb{Q}_i\mb{W})\ge \nu,\forall i\in \mc{N},\: Tr(\mb{W})=1,\:\mb{W} \succeq \mb{0}
\end{align}	
\end{subequations}
\normalsize
Here, $Tr()$ indicates the trace of the matrix. By solving the semidefinite problem \eqref{opt:sdp}, we will obtain the optimal $\mb{W}^*$ and $\nu_{sdp}$ (optimal $\nu$ value). The corresponding optimal vector $\mb{w}_{sdp}$ can be obtained by using eigenvalue decomposition of $\mb{W}^*$ or applying randomization technique \cite{luo2010semidefinite}.

\begin{figure}[h!]
	\vspace{-0.4cm}
	\centering
	\resizebox{7cm}{1.5cm}{
		\begin{tikzpicture}[->,>=stealth',scale=0.3]
		\node[state,text width=3.5cm,] (wblock)
		{Calculate $\mathbf{w}^*$, $\mu^*$ ($\nu_{min}$, $\nu_{max}$, $\nu_{sdp}$, $\nu_{gs}$) at the base station };
		\node[state,text width=5cm,right of=wblock,node distance=6cm,yshift=-0.75cm,xshift=0cm] (pblock)
		{Calculate $P^*$ at the base station};
		\node[state,text width=5.5cm,right of=wblock,node distance=6cm,yshift=0.75cm,xshift=0cm](rhoblock)
		{Calculate the $\rho^*$ at the power beacon};
		\draw[->] (wblock) -| (pblock);
		\draw[->] (wblock) -| (rhoblock);
		\end{tikzpicture} }
	\caption{\small{Block diagram of Stackelberg game equilibrium evaluation.}}
	\label{fig:blokdiag}
	\vspace{-0.2cm}
\end{figure}
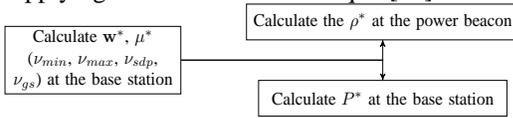

\begin{figure}[!h]
	\centering
	\includegraphics[scale=0.4]{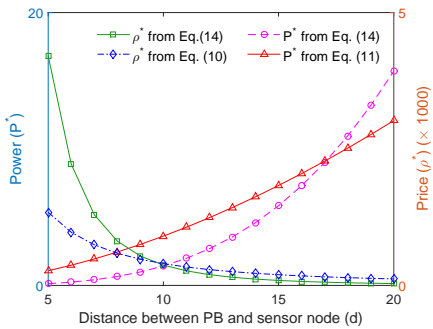}
	\caption{Plot of optimal power and price with respect to distance ($d$) between single antenna power beacon and a single sensor node for the utility function \eqref{eq:main_BS_util} and \eqref{eq:utilbssingle} for $M=1$, $\alpha=10^3,\;\beta=1,\;c=1$, $\theta^2=0.1$ and $\sigma^2=10^{-8}$. This plot compares the optimal solutions calculated using two different approaches: (a) solutions to Eq. \eqref{eq:optrho} and \eqref{eq:optpow} when $M=1$, (b) solutions to the system of equations \eqref{eq:prpnchi2}.}
	\label{fig:power_price_vs_distM1}
\end{figure}

\begin{figure}[!h]
	\centering
	\includegraphics[scale=0.38]{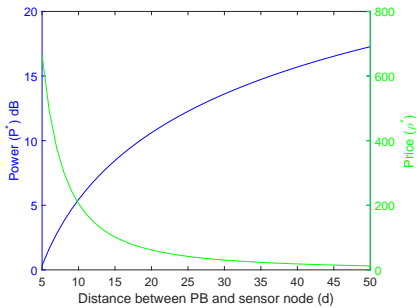}
	\caption{\small Plot of optimal power and price with respect to distance ($d$) between power beacon and a single sensor node. $\alpha=10^3,\;\beta=1,\;c=1$, $M=5$ and $\sigma=10^{-8}$.}
	\label{fig:power_price_vs_dist}
\end{figure}

\begin{figure*}[t!]
  \begin{minipage}{0.32\textwidth}
  	\centering
  	\includegraphics[scale=0.4]{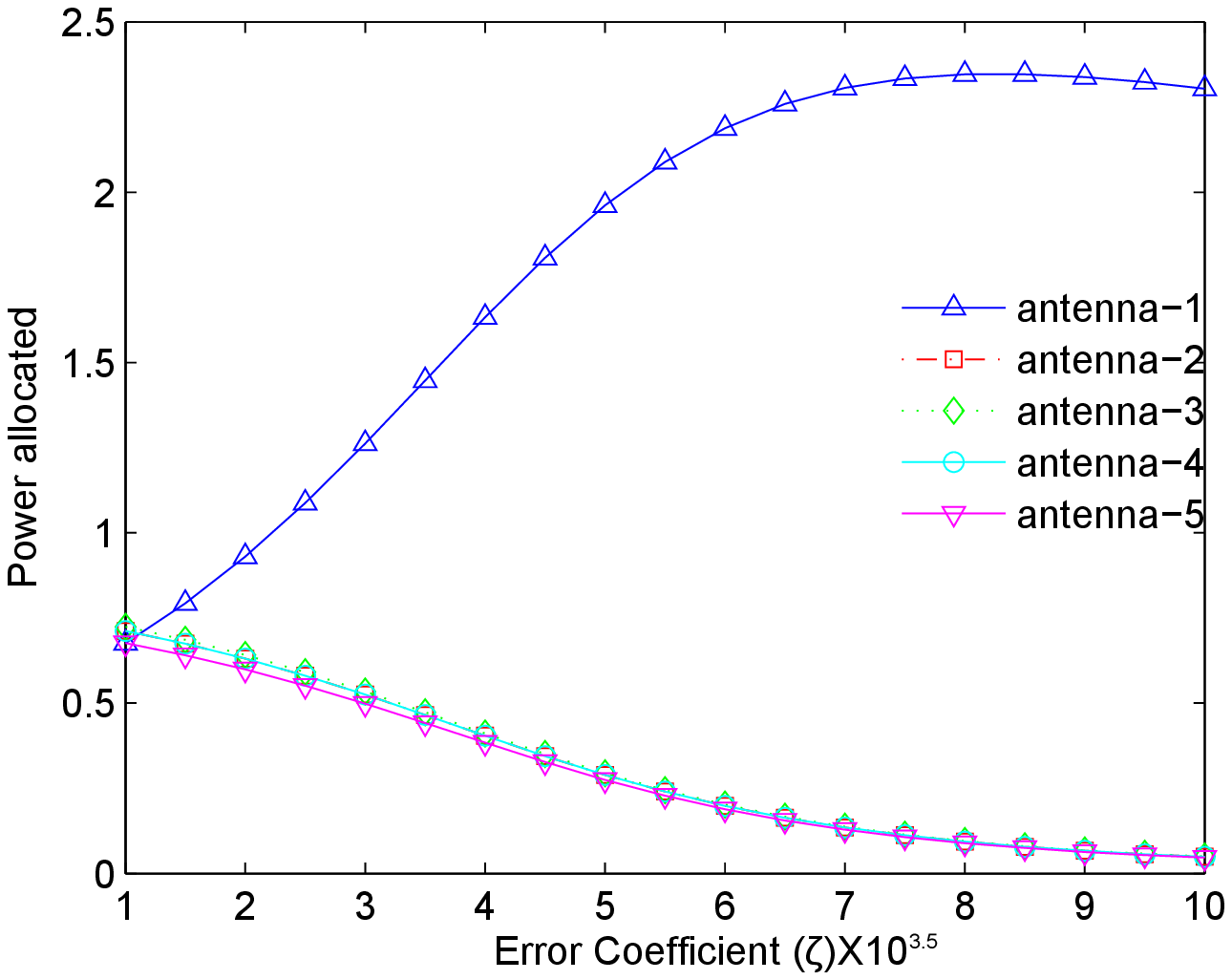}
  		\caption{\small Power allocated vs. error coefficient ($\zeta$) for a single sensor node. $M=5$ and $d=10$.}
  		\label{fig:error_vs_ant_power}
  \end{minipage}~~
  \begin{minipage}{0.64\textwidth}
    \centering
    \subcaptionbox{Power ($P$) vs. $d$.\label{fig:basestation_util_dist}}
    {\includegraphics[scale=0.4]{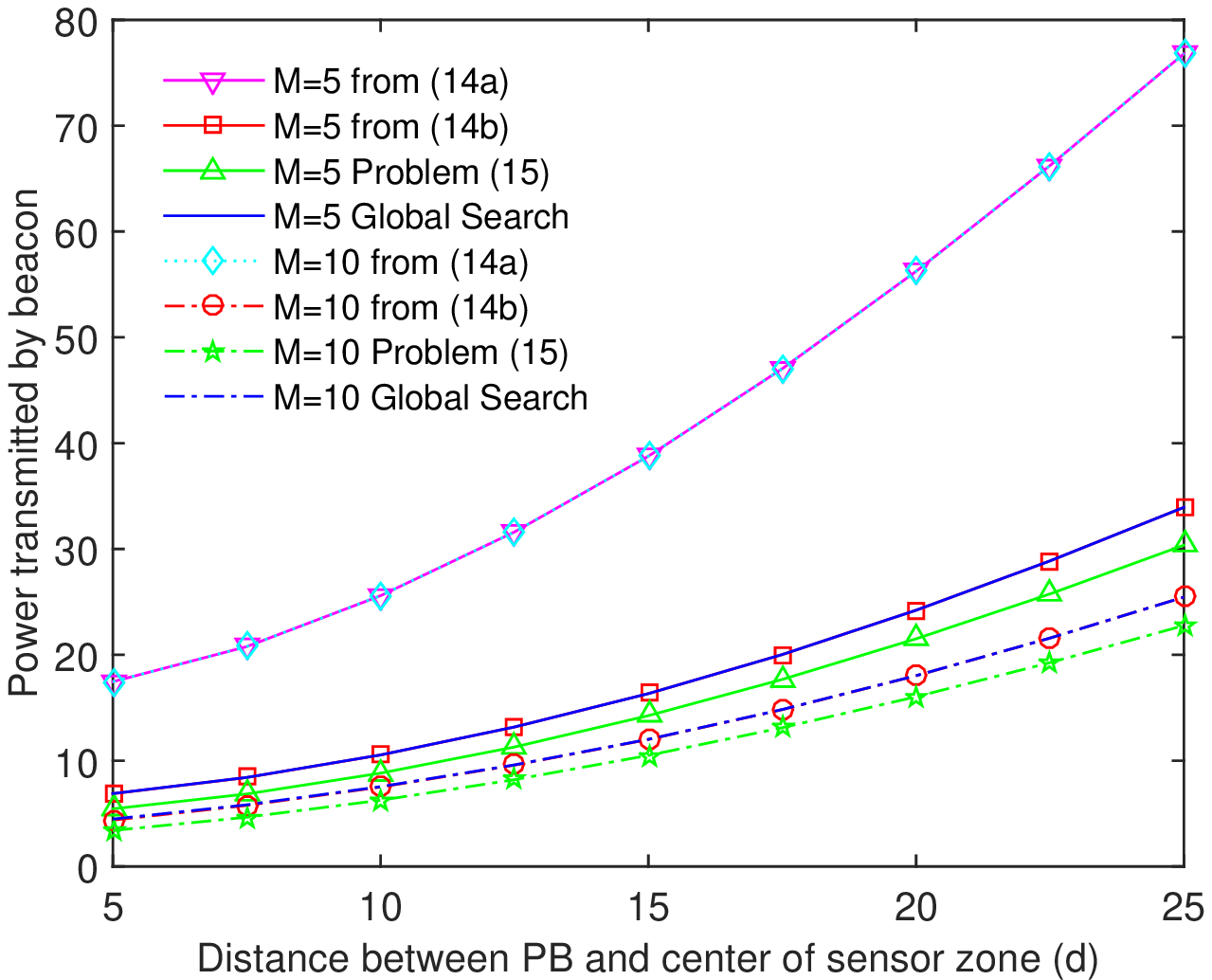}}~
    \subcaptionbox{Price ($\rho$) vs. $d$.\label{fig:fig_pwrbeacon_util_dist}}
     {\includegraphics[scale=0.4]{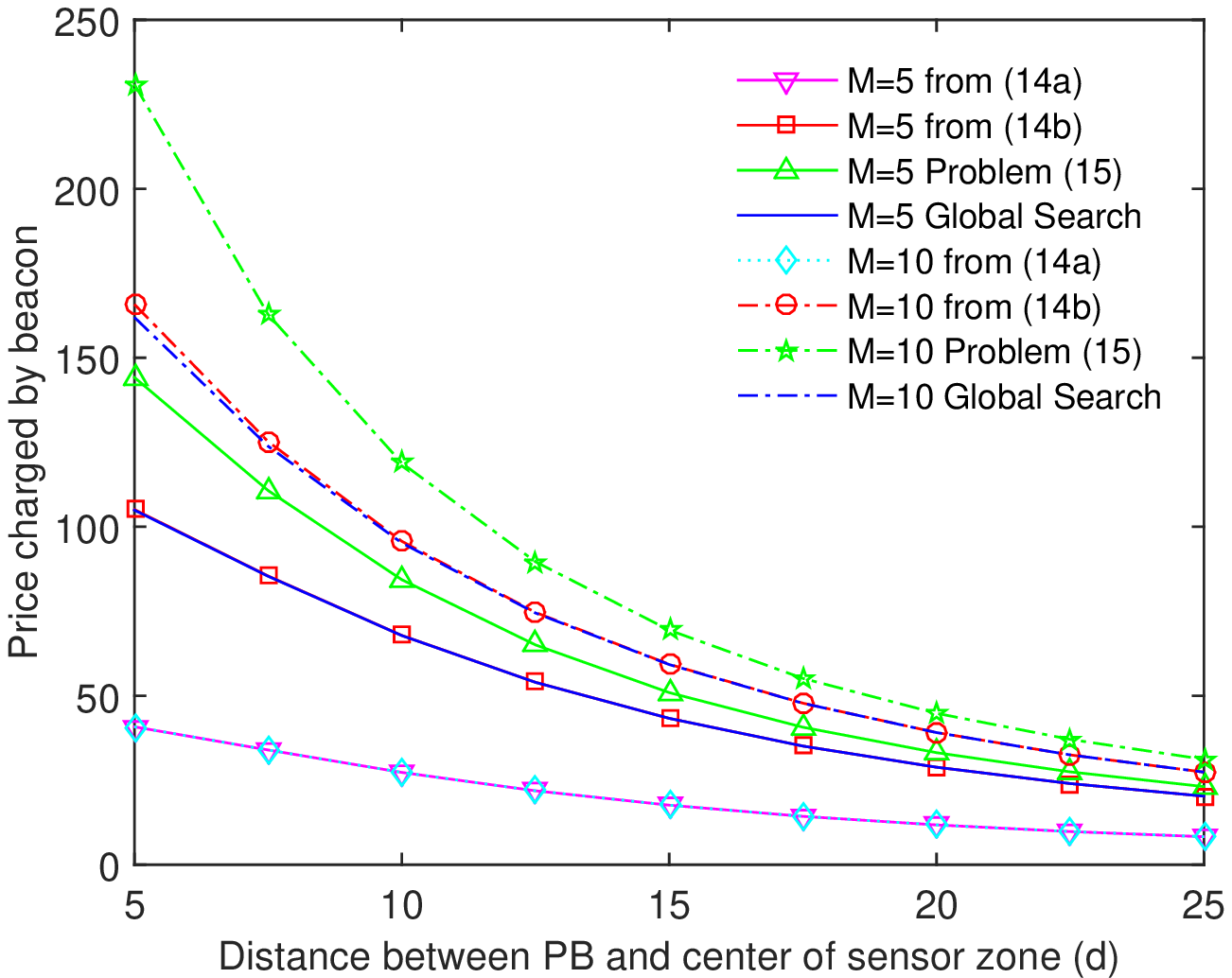}}
    \caption{\small  Power and price vs. $d$ for $M=5,\:10$. Plots corresponds to $\nu$ from \eqref{eq:lowbnds}, \eqref{eq:upbnds}, Problem \eqref{opt:sdp} and Global (exhaustive) search, respectively. $N=20$. Note: $\nu$ from global search and $\nu_{max}$ lead to almost same results.}
    \label{fig:power_price_vs_d}
  \end{minipage}
  \vspace{-0.2cm}
\end{figure*}

Fig. \ref{fig:blokdiag} summarizes the steps involved in evaluating the equilibrium for the proposed Stackelberg game.
\vspace{-0.3cm}
\section{Results}
\vspace{-0.1cm}

In this section, we have evaluated the Stackelberg equilibrium for different scenarios: (i) Single antenna power beacon, single sensor node (Fig. \ref{fig:power_price_vs_distM1}), 
(ii) Multiple antenna power beacon, single sensor node (Fig. \ref{fig:power_price_vs_dist}, \ref{fig:error_vs_ant_power}), (iii) Multiple antenna power beacon and multiple sensor nodes (Fig. \ref{fig:basestation_util_dist}, \ref{fig:fig_pwrbeacon_util_dist}). 

In Fig. \ref{fig:power_price_vs_dist}, we have plotted the optimal power and corresponding price  with respect to distance $d$ between the power beacon and the sensor node calculated using two approaches---(a) Eq. \eqref{eq:optrho}, \eqref{eq:optpow} and (b) Proposition \ref{prpn:chi2} (Eq. \eqref{eq:prpnchi2}).

In Fig. \ref{fig:power_price_vs_dist}, we have plotted the optimal power and corresponding price  with respect to distance $d$ between the power beacon and the sensor node calculated using \eqref{eq:optrho} and \eqref{eq:optpow}. Channel gain values are calculated based on the distance between the nodes using the following expression $h_{(.)}=d_{(.)}^{-\gamma/2}$, with a path-loss factor $\gamma=3.5$.  The ambient noise variance $\sigma^2=10^{-8}$. We have set the proportion of time for which energy is harvested $\tau=1/2$,  throughput threshold $\beta=1$, weighting parameter $\alpha=10^3$ and power beacon's operational cost parameter $c=1$. The error covariance matrix $\mathbf{\Sigma}$ is considered to be $\mathbf{I}/d^{\gamma}$, where $\mathbf{I}$ is the identity matrix. 

Fig. \ref{fig:error_vs_ant_power} studies the asymmetry in power allocation in antennas due to asymmetry in channel uncertainties for a single sensor node scenario (Sec. \ref{sec.singsensenode}). Without loss of generality, let $\zeta$ be the element at position $(1,1)$ of the diagonal covariance matrix $\mb{\Sigma}$, with other diagonal entries unchanged. We have plotted the power allocated to the $M$ antennas with respect to this uncertainty coefficient $\zeta$. We have evaluated the corresponding eigenvector $\mb{w}^*$, and calculated the power used by those antennas. It is evident that to mitigate the uncertainty, the power beacon will allocate more power to the first antenna, the channel corresponding to which has higher uncertainty (Fig. \ref{fig:error_vs_ant_power}).

In Fig. \ref{fig:basestation_util_dist} and \ref{fig:fig_pwrbeacon_util_dist}---corresponding to the multiple sensor node scenario (Sec \ref{sec.multsensnode})---we have plotted the power transmitted and the price charged by the power beacon with respect to the distance $d$ for different $M$ (number of antennas of the power beacon) values. The parameter $d$ is the distance between the center of the area where sensors are randomly placed, and the power beacon. Various curves correspond to $\nu_{min}$, $\nu_{max}$ (calculated from \eqref{eq:bounds}), the optimal $\nu$ calculated from SDP (Problem \eqref{opt:sdp}), and $\nu$ obtained from global (exhaustive) search. The number of sensor nodes is $N=20$. The base station coordinate is fixed at $(-10,0)$ and the power beacon coordinate varies as $(d,0)$.  Sensor nodes are randomly placed in a rectangular area with corners $(-4,-10),~(4,-10),~(4,10)$ and $(-4,10)$. The results are averaged over $100$ random positions of the sensors in this rectangular region. 

We find the upper bound in \eqref{eq:upbnds} to be tight, i.e., using the $\nu_{max}$ in \eqref{eq:multisensorbsutil}, we get a problem whose optimal solution ($P$, $\rho$) is almost exactly the same as the optimal obtained by global search. Also, for each $d$, the difference between the power (price) values calculated from the upper bound and exhaustive search is smaller than the difference between the power (price) values calculated from the SDP \eqref{opt:sdp} and exhaustive search. Note that, as the number of antennas $M$ increases, the power transmitted by the beacon decreases. Also, as expected, transmit power increases with distance. As $\wh{\mb{h}}_i\wh{\mb{h}}_i^\dagger$ is a rank-one matrix and $\mb{\Sigma}$ is a scaled identity matrix, the smallest eigenvalue of matrix $\big(\wh{\mb{h}}_i\wh{\mb{h}}_i^\dagger+\mb{\Sigma}\big)$ is independent of $\wh{\mb{h}}_i$ and depends only on the eigenvalues of $\mb{\Sigma}$, which is same for all $i$. Hence, from \eqref{eq:lowbnds} the power and price values corresponding to $\nu_{min}$ are overlapping.

\vspace{-0.2cm}
\section{Conclusion} 
\vspace{-0.1cm}
In this work, we have considered a Stackelberg game between the base station and the multi-antenna power beacon for wireless energy harvesting in a multiple sensor node network. We consider imperfect CSI between the power beacon and sensor nodes. The base station's utility is based on throughput non-outage probability and expenditure due to purchase of power. For the single sensor node scenario, we evaluate the equilibrium analytically. Equilibrium strategies corresponding to the upper and lower bounds on the utility of the base station is evaluated for multiple sensor nodes. 

\end{document}